\documentclass[lettersize,journal]{IEEEtran}
\IEEEoverridecommandlockouts
\usepackage{cite}
\usepackage{amsmath,amssymb,amsfonts}
\usepackage{algorithmic}
\usepackage{graphicx}
\usepackage{subcaption}
\usepackage{textcomp}
\usepackage{xcolor}
\usepackage{url}
\def\BibTeX{{\rm B\kern-.05em{\sc i\kern-.025em b}\kern-.08em
    T\kern-.1667em\lower.7ex\hbox{E}\kern-.125emX}}

\usepackage{physics}
\usepackage{pifont}
\usepackage{amssymb, amsmath,amsfonts,bm,bbm,amsthm}
\usepackage[capitalise]{cleveref}
\usepackage{enumitem}

\newtheorem{theorem}{Theorem}

\newtheorem{exemp}{Example}
\newtheorem{definition}{Definition}
\crefname{definition}{Definition}{Definitions}

\crefname{remark}{Remark}{Remarks}

\newcommand{\opn}[1]{\operatorname{#1}}

\newcommand{\aspas}[1]{``#1''}
\renewcommand{\vu}[1]{\ensuremath{\hat{#1}}}

\allowdisplaybreaks

\newcommand{\opq}{\vu{q}}

\newcommand{\opp}{\vu{p}}

\begin{document}

\title{Distributional Transform Based Information Reconciliation
}

\author{{Micael Andrade Dias and Francisco Marcos de Assis}

\thanks{Micael A. Dias and Francisco M. de Assis are with the Institute for Studies in Quantum Computing and Information Federal University of Campina Grande, Paraíba, Brazil (e-mails: micael.souza@ee.ufcg.edu.br and fmarcos@dee.ufcg.edu.br).}
\thanks{This work was supported in part by the National Council for Scientific and Technological Development (CNPq) under research Grant No. 305918/2019-2 and the Coordination of Superior Level Staff Improvement (CAPES/PROEX).}}

\maketitle

\begin{abstract}
    In this paper, we present an information reconciliation protocol designed for Continuous-Variable QKD using the Distributional Transform. By combining tools from copula and information theory, we present a method for extracting independent symmetric Bernoulli bits for Gaussian-modulated CVQKD protocols, which we called the Distributional Transform Expansion (DTE). We derived the expressions for the maximum reconciliation efficiency for both homodyne and heterodyne measurements, which, for the last, is achievable with an efficiency greater than 0.9 at a signal-to-noise ratio lower than -3.6 dB.
	
\end{abstract}

\begin{IEEEkeywords}
Information Reconciliation, Distributional Transform, CVQKD.
\end{IEEEkeywords}

\section{Introduction}

Quantum key distribution protocols perform the task of transmitting secret keys using quantum systems such that, in the end, Alice and Bob share identical binary sequences unknown to any other third part \cite{assche2006}. A broad family of QKD protocols uses continuous variable quantum systems to encode the secret key, called CVQKD protocols \cite{grosshans2002,grosshans2003,weedbrook2004,weedbrook2012,laudenbach2018, jouguet2011,jouguet2012}. A CVQKD protocol with Gaussian modulation of coherent states results in Alice and Bob sharing a pair of correlated sequences of Gaussian random variables. 

In the standard GG02 protocol \cite{grosshans2002}, Alice prepares a coherent state $\ket{\alpha_i}$, where $\alpha_i = q_i+jp_i$ comes from realizations of i.i.d. random variables $Q\sim P \sim\mathcal{N}(0,\tilde{V}_m)$. She sends it to Bob through a quantum channel, and at reception, Bob will perform homodyne detection by randomly switching between the quadratures. After $N$ rounds, Alice and Bob keep the matching values, owning the sequences $X_N = x_1,\cdots,x_N$ and $Y_N = x_1,\cdots,y_N$, respectively. A random subset of length $m<<N$ of both sequences is used to estimate the channel parameters, and the remaining sequences $X_n = X_N\setminus X_{[m]}$ and $Y_n = Y_N\setminus Y_{[m]}$ are called the \emph{raw key}.

	
Once the final secret key must be binary, the raw key values must be quantized and further corrected, both procedures that constitute the information reconciliation (IR) protocol \cite{assche2006,brassard1994,yan2008,nguyen2004,qian2009,lu2010}. This is in fact a crucial step for distilling secret keys and may be realized in a direct reconciliation (DR) direction, meaning that Bob must correct his binary sequences to match Alice's, or in a reverse reconciliation (RR), when Alice's is the one correcting her sequences towards Bob's key \cite{grosshans2003a}. Despite counterintuitiveness, reverse reconciliation is preferable as it allows the QKD protocol to run beyond the 3 dB loss limit of DR. 

Two widely used reconciliation protocols propose different ways to perform quantization. One is the Sliced Error Correction (SEC) protocol \cite{assche2004,jouguet2014}, which consists of a set of slicing functions for Alice and a set of estimators on Bob's side. After the slicing procedure has taken place, each emerging binary symmetric channel (BSC) can be treated separately with multilevel coding and multistage decoding (MLC-MSD), applying LDPC codes to perform error correction close to channel capacity \cite{jouguet2014,mani2021,bai2017}. The efficiency of the protocol depends not only on the error correction codes, but also on the quantization efficiency. However, the overall efficiency has been shown to lie above 0.9, specifically in the interval $1\sim3\opn{dB}$ of SNR (signal to noise ratio).

Another widely used method is multidimensional (MD) reconciliation, which applies $d$-dimensional rotations to simulate virtual channels close to BIAWGNC (Binary input AWGN channel) \cite{leverrier2008,jouguet2011,milicevic2018,mani2021}. This means that $d$ uses of the physical channel are assigned to $d$ approximate copies of a virtual BIAWGNC. Again, LDPC codes are used and MD reconciliation shows a high reconciliation efficiency for SNR around $0.5\opn{dB}$.

It is clear that the design of good reconciliation protocols for the low SNR regime is critical for CVQKD operation at long distances \cite{diamanti2016}. Here, we present an alternative method for extracting binary sequences from continuous-valued raw keys based on arguments from copula and information theories. More specifically, we extend the method presented in \cite{araujo2018}, which uses the distributional transform of continuous random variables (which is the principle of arithmetic source coding) to map the raw keys into the unit interval with uniform distribution. The bit sequences are then extracted with a simple binary expansion. We call this technique the Distributional Transform Expansion, the DTE. 

In contrast with SEC and MD reconciliation, the process of distilling bit sequences with DTE does not use any estimator or rotations in high-dimensional algebraic structures prior to the usage of error-correcting codes. In fact, a DTE reconciliation-based protocol has an analogous structure of SEC, that is, it allows for MLC-MSC, for example, but, as the results show, its best performance lies at very low signal-to-noise ratio, typically below -3.6 dB.

The paper is structured as follows. \Cref{sec:dte} defines the expansion of the distributional transformation and its application to the reconciliation problem. Its properties are explored in \Cref{sec:sub-chan-cap} by analyzing the subchannels induced by the binary expansions. \Cref{sec:recon-eff} develops its reconciliation efficiency and presents the main results. We conclude at \Cref{sec:conclusion} with the final considerations.

\section{Distributional Transform Expansion}\label{sec:dte}

Information reconciliation protocols aim to produce identical binary keys for both Alice and Bob with high probability using the data coming from the measurement outcomes of quantum communication, the raw key. Although it is possible to perform corrections on the continuous-valued data [ref shannon], there is no much practicality on this strategy. Then, the raw key must be quantized on at least one side (depending on whether DR or RR is performed) and the resulting bit sequences give rise to virtual classical channels modeling the correlations between Alice and Bob's strings.

The main approach to this problem are the SEC and MD reconciliation procedures, which present a way to extract bit sequences from continuous-valued data so that an error correction code could be applied, typically an LDPC code \cite{milicevic2018,mani2021,bloch2006,lodewyck2007}. The SEC protocol performs partitions on the real line (Alice's side) in order to assign bit sequences to each interval, and estimators are designed to recover such sequences on Bob's side. The resulting sequences are treated as bits transmitter through a binary symmetric channel. The MD reconciliation performs rotations such that the rotated values \aspas{looks like} are the result of transmitting bit sequences through a BIAWGN channel.

A relatively recent alternative, proposed by Araújo and Assis, proposes a different approach \cite{araujo2018}. It is based on two fundamental results of information theory and copula theory, which can be used to extract independent bit sequences from numbers lying in the unit interval. In the following, we present the definition of a generalized inverse of a distribution function and then the result affirming its uniform distribution in the unit interval.


\begin{definition}
    Let $F: \mathbb{R}\rightarrow\mathbb{I}$ be a distribution function. The quasi-inverse of $F$, also known as the generalized inverse, is the function $F^{(-1)}:\mathbb{I}\rightarrow\mathbb{R}$ given by
    \begin{equation}
        F^{(-1)} = \inf\{x\in\mathbb{R}: F(x)\geq t\},\,t\in(0,1],
    \end{equation}
    \noindent where $F^{(-1)}(0) = \inf\{x\in\mathbb{R}: F(x)> 0\}$.
\end{definition}

\begin{theorem}[\cite{sempi2016}]\label{th:distrib-transform}
    Let $X$ be a random variable with distribution function $F_X$ and $F_X^{(-1)}$ its \textit{quasi-inverse}. Then	
    \begin{enumerate}
        \item If $F$ is continuous, then $U = F_X(X)$ is uniformly distributed on $[0,1]$.
        \item If $U$ is a uniformly distributed random variable in $[0,1]$, then $Y = F_X^{(-1)}(U)$ has a distribution function according to $F_X$.
    \end{enumerate}
\end{theorem}

The transformation mentioned in the first part of \Cref{th:distrib-transform} is known as the Distributional Transform and ensures that transforming a random variable by its continuous distribution function always leads to a uniform distribution in the unit interval. Together with the fact that the bits in the binary expansion of a random variable with uniform distribution on $[0,1]$ are independent and Bernoulli$(\frac{1}{2})$ \cite{cover2006}, one can use the distributional transform to map the raw key values on the unit interval and apply a binary expansion on the resulting value. 

\begin{figure}[!tb]
    \centering
    \includegraphics[width=\columnwidth]{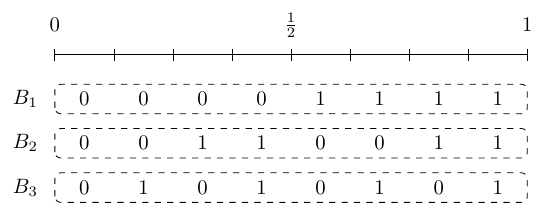}
    \caption{\label{fig:bin-exp-exaple}Unit interval partition according to a $3$-bit binary expansion and the bits corresponding values.}
\end{figure}

The number $d\in\qty[0,1]$ can be expanded in the binary basis with $l$ bit precision according to the following rule,
\begin{align}\label{eq:bin-exp}
d \mapsto 0.b_1b_2\cdots b_l, & & \sum_{i=1}^{l-1}b_i\frac{1}{2^i}\leq d \leq \sum_{i=1}^{l-1}b_i\frac{1}{2^i} + \frac{1}{2^l},
\end{align}
\noindent and we call $\bm{b} = b_1b_2\cdots b_l$ the corresponding bit sequence. Each bit has information about where the real number $d$ lies in the unit interval: the first bit ($b_1$) announces if $d\in [0,\frac{1}{2})$ or $d\in[\frac{1}{2},1]$, the second one ($b_2$) informs if $d$ lies in the left or right quarter of the $1/2$ interval indicated by $b_1$, that is, if $d\in [0,\frac{1}{4})$ or $d\in[\frac{1}{4},\frac{1}{2}]$ given $b_1=0$ or $b_1=1$, respectively, and so on. In the \Cref{fig:bin-exp-exaple} it is depicted the bit values for each interval in a 3-bit expansion.

This procedure for extracting independent equiprobable bits from realizations of a continuous-valued random variable $X$ can be formalized as what we call the distributional transform expansion of $X$.

\begin{definition}\label{def:DTE}
    Let $X$ be a random variable with a continuous distribution function $F_X$ and $\mathcal{Q}:\qty[0,1]\mapsto\qty{0,1}^l$ a function that gives a binary expansion as in \Cref{eq:bin-exp}. The Distributional Transform Expansion (DTE) is defined as
    \begin{equation}
        \mathcal{D}(X) = \mathcal{Q}(F_X(X)).
    \end{equation}
    Once the bits in the binary expansion are independent, it is possible to factor $\mathcal{D}(X) = \mathcal{D}_1(X)\cdots\mathcal{D}_l(X)$, where $\mathcal{D}_i(X) = \mathcal{Q}_i(F_X(X))$ is the function $\mathcal{Q}_i:\qty[0,1]\mapsto\qty{0,1}$ that computes the $i$-th bit in \Cref{eq:bin-exp} with the property $\mathcal{D}_i(X)\sim\opn{Bern}(\frac{1}{2})$. We call by l-$\mathcal{D}(X)$ the DTE expansion of $F$ with length $l$.
\end{definition}

Alice and Bob can use the DTE to produce binary sequences from their continuous-valued data: 
\begin{enumerate}
	\item Alice and Bob have the sequences of Gaussian variables $X = X_1,\cdots,X_n$ and $Y = Y_1,\cdots,Y_n$ after quantum communication and parameter estimation;
	\item Alice [Bob in RR] compute $\mathcal{D}(X) = (\mathcal{D}_1(X), \cdots, \mathcal{D}_l(X))^T$ for each raw key element [$\mathcal{D}(Y) = (\mathcal{D}_1(Y), \cdots, \mathcal{D}_l(Y))^T$, in RR]. The resulting bit sequences can be expressed as matrices,
	\begin{align}\label{eq:expansion}
	X & \mapsto\mqty[\mathcal{D}_1(X_1) & \cdots & \mathcal{D}_1(X_2)\\
	\mathcal{D}_2(X_1) & \cdots & \mathcal{D}_2(X_2)\\
	\vdots             & 		 & \vdots \\
	\mathcal{D}_l(X_1) & \cdots & \mathcal{D}_l(X_2)\\ ] = \mqty[\mathcal{D}_1(X) \\ \mathcal{D}_l(X) \\ \vdots \\ \mathcal{D}_l(X)],\\
	Y & \mapsto\mqty[\mathcal{D}_1(Y_1) & \cdots & \mathcal{D}_1(Y_2)\\
	\mathcal{D}_2(Y_1) & \cdots & \mathcal{D}_2(Y_2)\\
	\vdots             & 		 & \vdots \\
	\mathcal{D}_l(Y_1) & \cdots & \mathcal{D}_l(Y_2)\\ ] = \mqty[\mathcal{D}_1(Y) \\ \mathcal{D}_l(Y) \\ \vdots \\ \mathcal{D}_l(Y)],
	\end{align} 
	\item Each one of the $l$ pairs of sequences $(\mathcal{D}_i(X),Y)$ [$(\mathcal{D}_i(Y),X)$ in RR] can be interpreted as a Binary-Input AWGN channel 
	and Bob [Alice in RR] can retrieve Alice's [Bob in RR] binary sequences by using an error correcting code.
\end{enumerate}

\begin{exemp}
	Let $X \sim\mathcal{N}(0,1)$, $Z \sim\mathcal{N}(0,0.5)$ with $X\perp Z$ and $Y = X+Z$. Assume the realizations $x = \qty{0.491,  0.327, -0.652 , -1.096, -0.023}$ and $z = \qty{-0.722,  0.942,  0.191,  0.198, -0.370}$. Then
	\begin{align*}
	F_X(x) &= (0.688, 0.628, 0.257, 0.136, 0.491) \\ 
               &\mapsto\mqty[1 & 1 & 0 & 0 & 0 \\ 
                             0 & 0 & 1 & 0 & 1 \\ 
                             1 & 1 & 0 & 1 & 1 \\
                             1 & 0 & 0 & 0 & 1 \\
                             0 & 0 & 0 & 0 & 1]\\
	F_Y(y) &= (0.425, 0.850, 0.353, 0.231, 0.374) \\
               &\mapsto\mqty[0 & 1 & 0 & 0 & 0 \\ 
                             1 & 1 & 1 & 0 & 1 \\ 
                             1 & 0 & 0 & 1 & 0 \\
                             0 & 1 & 1 & 1 & 1\\
                             1 & 1 & 1 & 1 & 1].
	\end{align*}
\end{exemp}

As the bits in the expansion are pairwise independent, it is also possible that Alice and Bob perform the DTE on their sequences and treat the errors between $\mathcal{D}_i(X)$ and $\mathcal{D}_i(Y)$ as transmitted over a binary symmetric channel (BSC) with transition probability $p_i$. This approach was used in \cite{araujo2018} where they showed that reconciliation can be obtained in the first two subchannels with $4\cdot 10^4$ sized LDPC codes in at most 40 decoding iterations with 4.5 dB SNR. However, the analysis was restricted to CVQKD protocols with homodyne detection, and reconciliation efficiency was not addressed. 

Those two possible approaches, error correction over the BSC and BIAWGN induced channels, are the ones that intuitively appears after performing DTE on the raw key sequences $X$ and $Y$. Clearly, the BSC approach must not have a better performance than BIAWGN due to the data processing inequality that ensures that $I(\mathcal{D}_i(X);Y)\leq I(\mathcal{D}_i(X),\mathcal{D}_i(Y))$. The next section will focus on characterizing those two kinds of subchannel and providing an upper bound on the reconciliation efficiency.

\subsection{Impracticality of Bivariate DTE}
    
    The DTE defined as in \Cref{def:DTE} uses the univariate distributional transform to extract independent binary sequences from continuous valued data. Then, one could reasonably ask: what about a bivariate distributional transform such as $V = F_{QP}(Q,P)$? This goes back to CVQKD protocols with heterodyne measurement, where both quadratures modulation and detection outcomes are used to distill a secret key. It turns out that the Kendall distribution function \cite{nelsen2003} of a random vector $X = X_1,\cdots, X_d$ with joint distribution $F$ and marginals $F_1,\cdots, F_d$ defined as $\kappa_F = \opn{Pr}\qty{F(X_1,\cdots, X_d)\leq t}$ does not need to be uniform in $[0,1]$ \cite[Definition 3.9.5]{sempi2016}. In fact, for the bivariate case of independent random variables, $\kappa_F$ is not uniform, which is exactly the case of heterodyne measured CVQKD protocols and the DTE reconciliation would not work.

\section{DTE Sub-channels Capacities}\label{sec:sub-chan-cap}

Given that Alice and Bob can use the DTE to extract binary sequences from the continuous valued raw keys and those binary sequences can behave as a BSC or BIAWGN depending on whether the DTE is performed only on $X$, $Y$ or both, it is necessary to estimate those BIAWGN and the BSC's subchannel capacities. This will allow one to obtain an upper bound to reconciliation efficiency. For BSC's, the transition probabilities $p_i = \opn{Pr}\qty{\mathcal{D}_i(X) \neq \mathcal{D}_i(Y)}$ must be obtained, which is the approach in \cite{araujo2018}. The BIAWGN capacities are more involved and require estimating $I(\mathcal{D}_i(X);Y)$ for DR and $I(\mathcal{D}_i(Y);X)$ for RR.

In the following, the induced AWGN channel connecting the classical random variables $X$ of Alice's modulation and $Y$ of Bob's measurement outputs, whose noise appears as a function of the quantum channel parameters. Expressions for reconciliation efficiency are also given for both direct and reverse reconciliation.


\subsection{Equivalent AWGN Channel}

Starting with a Gaussian modulated protocol with homodyne detection (the GG02 \cite{grosshans2002}), in the EB protocol version, Alice and Bob's shared state after the quantum channel transmission and prior the detection has the following covariance matrix \cite{laudenbach2018},
\begin{equation}
\bm\Sigma'_{AB} = \begin{pmatrix}
V\bm{I}_2 & \sqrt{\tau}\sqrt{V^2-1}\bm{Z} \\ \sqrt{\tau}\sqrt{V^2-1}\bm{Z} & \qty[\tau V_m + 1 + \xi]\bm{I}_2
\end{pmatrix},
\end{equation}
\noindent where $V = V(\opq) = V(\opp) = V_m+1$ is the total quadrature variance, $V_m = 4\tilde{V}_m$ and $\xi = 2\bar{n}(1-\tau)$ is the channel excess noise from the thermal noise $\varepsilon = 2\bar{n}+1$, being $\bar{n}$ the mean thermal photons excited in the mode. Bob's mode is in a zero mean thermal state with $\bm\Sigma'_B = \qty[\tau V_m + 1 + \xi]\bm{I}_2$ and, when he homodynes, its output probability distribution is the Gaussian \cite{nha2005},
\begin{equation}
p_Y(y) = \sqrt{\frac{1}{2\pi\sigma_Y^2}}\exp(-\frac{1}{2}\frac{y^2}{\sigma_Y^2}),
\end{equation}
\noindent where we made $\sigma_Y^2 = (\tau V_m+\xi+1)/4$. Recalling that $X\sim\mathcal{N}(0,\tilde{V}_m)$, we can restate Bob's output as $Y = \sqrt{\tau}X+Z'$, with $Z'\sim\mathcal{N}\qty(0,\frac{\xi+1}{4})$ and $X\perp Z$. With a normalization, we get the AWGN channel model $Y = X+Z$, with $Z'/\sqrt{\tau} = Z \sim\mathcal{N}(0,\sigma_{Z_1}^2 = (\xi+1)/4\tau)$ and $\sigma_Y^2 = \tilde{V}_m + \frac{\xi+1}{4\tau}$. It yields the signal to noise ratio
\begin{equation}\label{eq:snr-hom}
\operatorname{SNR}_{hom} = \frac{\tau V_m}{1+\xi}.
\end{equation}

When Bob performs heterodyne (or double homodyne) detection, which is the case in the \textit{no-switching} protocol \cite{weedbrook2004}, his mode goes through a 50:50 beam spliter, the two resulting modes are described by the covariance matrix \cite{laudenbach2018},
\begin{equation}
\bm\Sigma'_{B_1B_2} = \begin{pmatrix}
\qty(\frac{\tau}{2}V_m + 1 + \frac{\xi}{2})\bm{I}_2 & -\frac{\tau V_m\xi}{2}\bm{I}_2\\\-\frac{\tau V_m\xi}{2}\bm{I}_2 & \qty(\frac{\tau}{2}V_m + 1 + \frac{\xi}{2})\bm{I}_2
\end{pmatrix},
\end{equation}
\noindent and each splitted mode is homodyned such that the $\opq/\opp$ quadrature measurements are equally distributed as $Y_q\sim Y_p\sim\mathcal{N}(0,\sigma_Y^2 = (\frac{\tau}{2}V_m + 1 + \frac{\xi}{2})/4)$ and they can be seen as $Y_* = \sqrt{\tau/2}X+Z'$ where $Z'\sim\mathcal{N}(0,\frac{1+\xi/2}{4})$. As in the homodyne case, it can be normalized and we get $Y_* = X+Z$ with $\sqrt{\frac{2}{\tau}}Z' = Z\sim\mathcal{N}(0,\sigma_{Z_2}^2=\frac{1+\xi/2}{2\tau})$ and $\sigma_Y^2 = \tilde{V}_m+(\xi/2+1)/2\tau$. The resulting SNR is then,
\begin{equation}\label{eq:snr-het}
\operatorname{SNR}_{het} = \frac{\frac{\tau}{2}V_m}{1+\frac{\xi}{2}}.
\end{equation}

It is important to note that for homodyne or heterodyne detection, the signal-to-noise ratio is a function of the modulation variance (known by Alice e Bob prior to the protocol execution) and the channel invariants ($\tau$ and $\xi$, both to be obtained by parameter estimation). Therefore, given the values of $\tilde{V}_m$, $\tau$ and $\xi$, $\operatorname{SNR}_{hom} \neq \operatorname{SNR}_{het}$. Also, given the symmetry in modulation and the independence between the quadratures, the homodyne and heterodyne reconciliation efficiencies can be estimated simply by simulating an AWGN channel with the appropriate noise variance. For the heterodyne measurement, it is sufficient to estimate only one quadrature measurement once both quadratures are statistically equivalent.

\subsection{DTE Sub-channels Capacities}

With the AWGN channels connecting $X$ and $Y$ set up, it is possible to simulate what Alice and Bob would have after exchanging coherent states and performing coherent measurement by randomly drawing Gaussian random variables. For the continuously valued raw keys, the N realization of $X\sim\mathcal{N}(0,\tilde{V}_m)$ corresponds to Alice's modulated states, as well as the N realizations of $Z \sim\mathcal{N}(0,\sigma_{Z_1}^2)$ or $Z\sim\mathcal{N}(0,\sigma_{Z_2}^2)$ to give Bob's output measurements $Y=X+Z$. Then, an l-DTE with $l=4$ is applied to $X$, $Y$ or both to estimate the subchannel parameters.

\begin{figure}[tb]
	\centering
	\includegraphics[page=10]{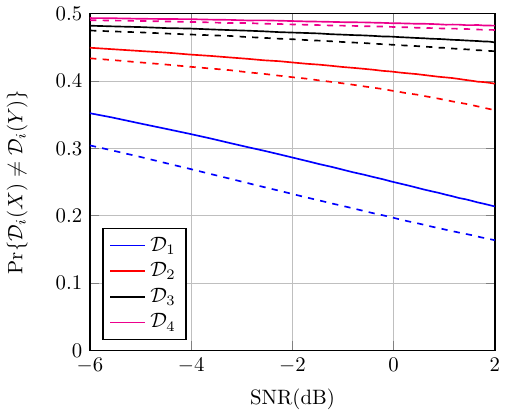}
	\caption{\label{fig:bit-error-probability}Transition probability in the four-bit DTE on both Alice and Bob's side in a Gaussian modulated CVQKD protocol. The probabilities were estimated drawing $N=10^4$ realizations of Alice's random variable and repeating the experiment $10^3$ times. The parameters were $\tilde{V}_m = 1$ and $\xi=0.02$ for both heterodyne (solid lines) and homodyne (dashed line) detection. 
	}
\end{figure}

\begin{figure*}[!t]
	\centering
	\begin{subfigure}{\columnwidth}
		\centering
		\includegraphics[page=5]{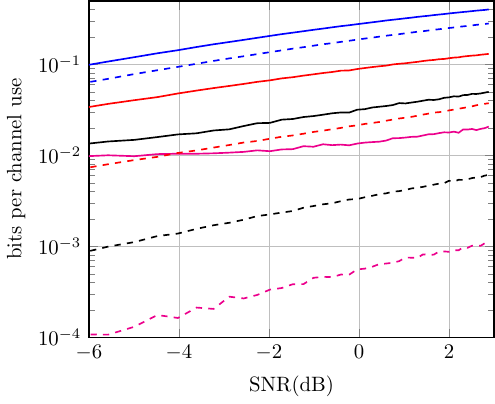}
		\caption{\label{fig:sub-chan-capacities-het-RR}Heterodyne detection, reverse reconciliation.}
	\end{subfigure}%
        \begin{subfigure}{\columnwidth}
		\centering
		\includegraphics[page=6]{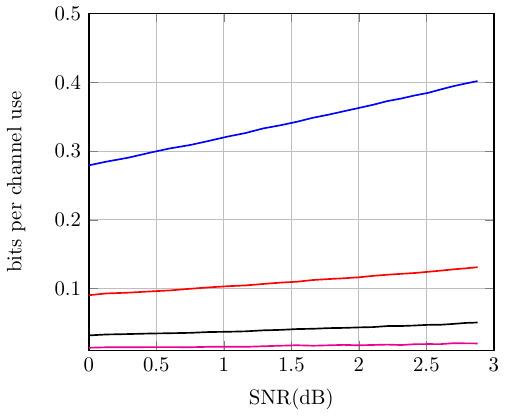}
		\caption{\label{fig:sub-chan-capacities-het--DR}Heterodyne detection, direct reconciliation.}
	\end{subfigure}

	\begin{subfigure}{\columnwidth}
		\centering
		\includegraphics[page=8]{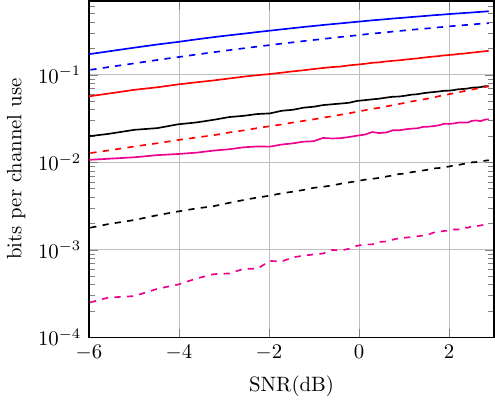}
		\caption{\label{fig:sub-chan-capacities-hom-RR}Homodyne detection, reverse reconciliation.}
	\end{subfigure}%
         \begin{subfigure}{\columnwidth}
            \centering
            \includegraphics[page=9]{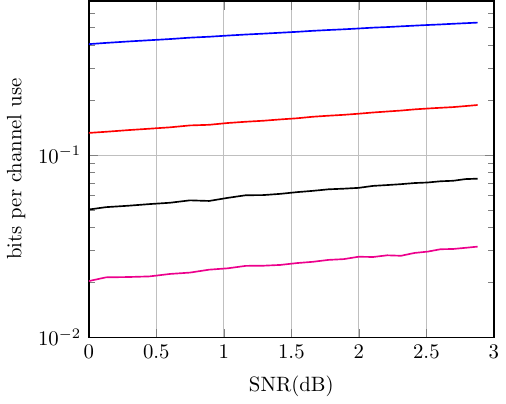}
            \caption{\label{fig:sub-chan-capacities-hom-DR}Homodyne detection, direct reconciliation.}
        \end{subfigure}
	\caption{\label{fig:sub-chan-capacities}Sub-channels capacities of the BIAWGN and BSC induced by the DTE in a Gaussian modulated CVQKD protocol with heterodyne/homodyne detection and direct/reverse reconciliation directions. Capacity was obtained by drawing $N=10^4$ realizations of Alice's random variable and estimating the mutual information $I(\mathcal{D}_i(Y);X)$ (solid lines) and $I(\mathcal{D}_i(X);Y)$ for BIAWGN, and computing $C_{BSC_i} = 1 - H(p_i)$ for BSC (dashed lines), where $p_i$ is the estimated transition probability shown in \Cref{fig:bit-error-probability}. The experiments were repeated $10^3$ times for both detection methods and the results presented are the mean values. In any plot (solid/dashed lines or cross dots), $\mathcal{D}_1$ is the at the top and $\mathcal{D}_4$ at the bottom.}
\end{figure*}

First, we characterize the BSC's subchannels by estimating the transition probabilities $p_i = \opn{Pr}\qty{\mathcal{D}_i(X) \neq \mathcal{D}_i(Y)}$. For the BIAWGNs, we used the entropy estimators available in \cite{steeg2016}, which implement Kraskov's mutual information estimator \cite{kraskov2004} to get $I(\mathcal{D}_i(X);Y)$ and $I(\mathcal{D}_i(Y);X)$. The results are plotted in \Cref{fig:bit-error-probability,fig:sub-chan-capacities}. It can be seen that as the expansion goes further on gathering bits from the continuous sequences $X$ and $Y$, the resulting subchannels becomes more noisy, easily approaching the behavior of a fair coin in \Cref{fig:bit-error-probability}. It is worth pointing out that the BSC's transition probabilities do not depend on the reconciliation direction, as well as its capacity. 

The subchannel capacities for BIAWGN and BSC are plotted in \Cref{fig:sub-chan-capacities} for both RR and DR with heterodyne and homodyne detection. First, the BSC's capacities (dashed lines in \Cref{fig:sub-chan-capacities-het} and \Cref{fig:sub-chan-capacities-hom-RR}) are far apart from the BIAWGN ones (solid lines), from which we conclude that applying the DTE on both Alice's and Bob's sequences will not result in a good reconciliation efficiency. The respective capacities for the BIAWGN channels when DR is considered are plotted in \Cref{fig:sub-chan-capacities-het--DR,fig:sub-chan-capacities-hom-DR}, which can be seen to be very close to the RR direction. Although DR is restricted to $\tau>0.5$ and, as will be seen in the next section, the best efficiency of DTE is found in the region with $\opn{SNR}<0\opn{dB}$. Then, further analysis on the reconciliation efficiency will be restricted to the RR direction.



\section{Reconciliation Efficiency}\label{sec:recon-eff}

The $l$ bit quantization process performed by the DTE is a function $\mathcal{D}:\mathbb{R}\mapsto\qty{0,1}^l$ that can be broken down as $l$ single-bit quantization functions $\mathcal{D}_i:\mathbb{R}\mapsto\qty{0,1}$, $i=1,\cdots,l$, as stated in \Cref{def:DTE}. Here, we derive the general expressions for the reachable reconciliation efficiencies when using the DTE to distill secret keys. In the following, we use the right and left arrows in the exponent to indicate direct and reverse reconciliation directions, respectively.

    \subsection{Direct reconciliation}
    
    First, consider that Alice applies the DTE to the $n$ realizations of her Gaussian variables $X$ so that Bob must recover her binary sequence. The secret rate \textit{per transmitted state} in direct reconciliation (DR) is given by \cite{assche2006},
    \begin{align}
        K^\rightarrow &= H(\mathcal{D}(X)) - \chi(X, E) - l^{-1}|M^\rightarrow|,\\
                      &= \beta^\rightarrow I(X;Y) - \chi(X, E),
    \end{align}
    \noindent where $\chi(X, E)$ is the Holevo bound on Eve's accessible information, being $E$ her \textit{ancilla} systems, $|M^\rightarrow|$ the amount of side information Alice must send to Bob in direct reconciliation, and
    \begin{equation}
        \beta^\rightarrow = \frac{H(\mathcal{D}(X)) - l^{-1}|M^\rightarrow|}{I(X;Y)}.
    \end{equation}
    
    The upper bound on the reconciliation efficiency is reached when Alice uses the minimum amount of side information, that is, when $|M|\cdot l^{-1} = H(\mathcal{D}(X)|Y)$ and the maximum reconciliation efficiency reads
    \begin{equation}\label{eq:max-recon-dir}
        \beta^\rightarrow_{max} = \frac{H(\mathcal{D}(X)) - H(\mathcal{D}(X)|Y)}{I(X;Y)} \geq \beta^\rightarrow.
    \end{equation}
    
    With a closer look at the conditional entropy in \Cref{eq:max-recon-dir}, one derives
    \begin{align}
        H(\mathcal{D}(X)|Y) &\overset{(a)}{=} H(\mathcal{D}_1(X), \cdots, \mathcal{D}_l(X)|Y),\\\nonumber
            &\overset{(b)}{=} H(\mathcal{D}_1(X)|Y) + H(\mathcal{D}_2(X)|\mathcal{D}_1(X),Y)+\\ 
            &\cdots + H(\mathcal{D}_l(X)|\mathcal{D}_{l-1}(X), \cdots, \mathcal{D}_1(X),Y)\\
            &\overset{(c)}{=} \sum_{i=1}^{l}H(\mathcal{D}_i(X)|Y))\\
            &\overset{(d)}{=} \sum_{i=1}^{l}\qty(H(\mathcal{D}_i(X)) - I(\mathcal{D}_i(X);Y))\\
            &\overset{(e)}{=} l - \sum_{i=1}^{l}I(\mathcal{D}_i(X);Y),
    \end{align}
    \noindent where (a) comes from \Cref{def:DTE}, (b) is the chain rule for the joint entropy, (c) is due to $\mathcal{D}_i(X)\perp\mathcal{D}_j(X),i\neq j$, (d) comes from the identity $H(A|B) = H(A) - I(A;B)$ and (e) follows from $\mathcal{D}_i(X)\sim\opn{Bern}(\frac{1}{2})$, which gives $H(\mathcal{D}_i(X)) = 1$. This concludes
    \begin{align}\label{eq:recon-eff-dir_}
        \beta^\rightarrow_{max} &= \frac{H(\mathcal{D}(X)) - l + \sum_{i=1}^{l}I(\mathcal{D}_i(X);Y)}{I(X;Y)},\\\label{eq:recon-eff-dir}
                                &= \frac{\sum_{i=1}^{l}I(\mathcal{D}_i(X);Y)}{I(X;Y)}.
    \end{align}
    \noindent once $H(\mathcal{D}(X)) = H(\mathcal{D}_1(X), \cdots, \mathcal{D}_l(X)) = H(\mathcal{D}_1(X))+\cdots+H(\mathcal{D}_l(X)) = l$. That is, the maximum efficiency is proportional to the fraction of mutual information in the subchannels that the DTE can extract from the actual AWGN channel.
    
    \subsection{Reverse reconciliation}
    
    In the case of reverse reconciliation, Bob is the one performing the DTE on his Gaussian sequence $Y$ and must send some side information to Alice so that she can recover his sequences. In this way, the secret key rate \textit{per transmitted state} in reverse reconciliation becomes
    \begin{align}
        K^\leftarrow &= H(\mathcal{D}(Y)) - \chi(Y, E) - l^{-1}|M^\leftarrow|,\\
                     &= \beta^\leftarrow I(X;Y) - \chi(Y, E),
    \end{align}
    \noindent and, analogously to the DR case, $\chi(Y, E)$ the Holevo bound on Eve's accessible information to Bob's system, $|M^\leftarrow|$ is the amount of side information Bob must send to Alice, and
    \begin{equation}
        \beta^\leftarrow = \frac{H(\mathcal{D}(Y)) - l^{-1}|M^\leftarrow|}{I(X;Y)}.
    \end{equation}
    
    Following the same procedure of direct reconciliation, when $l^{-1}|M^\leftarrow|\rightarrow H(\mathcal{D}(Y)|X)$, the maximum reconciliation efficiency in the reverse direction is given by
    \begin{align}\nonumber
        \beta^\leftarrow \leq \beta^\leftarrow_{max} &= \frac{H(\mathcal{D}(Y)) - H(\mathcal{D}(Y)|X)}{I(X;Y)},\\\label{eq:recon-eff-rev}
                                                     &= \frac{\sum_{i=1}^{l}I(\mathcal{D}_i(Y);X)}{I(X;Y)}.
    \end{align}
    
    \subsection{Some Comments on the Reconciliation Efficiency}
    
    Firstly, in both direct and reverse reconciliation, exchanging the minimum amount of side information implies that error correction codes must run at channels capacity, and this is the only factor that affects the efficiency of the protocol. The \Cref{eq:recon-eff-dir_} is the same as in several information reconciliation papers using SEC \cite{jouguet2014}. Although, the entropy\footnote{In this paragraph we use $\mathcal{Q}$ as a generic quantization function.} $H(\mathcal{Q}(X))$ in the SEC protocol does not necessarily equals to $|\mathcal{Q}(X)|$, and such equality comes naturally in the DTE due to the independency between its bits.

    \begin{figure}[!t]
	\centering
	\begin{subfigure}{\columnwidth}
		\centering
		\includegraphics[page=2]{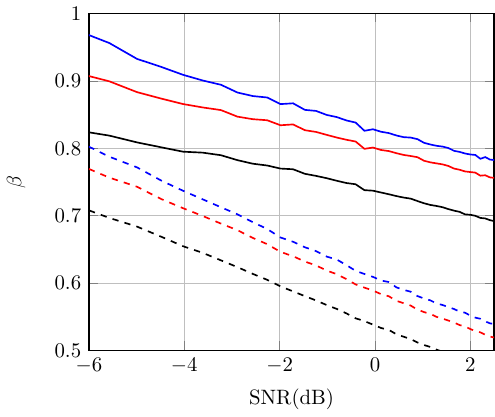}
		\caption{\label{fig:sub-chan-capacities-het}Reverse reconciliation.}
	\end{subfigure}

	\begin{subfigure}{\columnwidth}
		\centering
		\includegraphics[page=3]{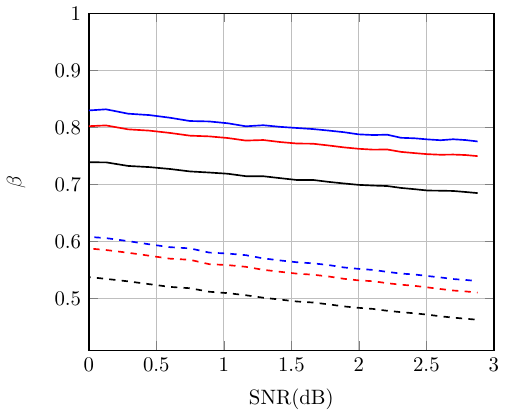}
		\caption{\label{fig:sub-chan-capacities-hom}Direct reconciliation.}
	\end{subfigure}
        \caption{\label{fig:DTE-recon-eff}Reconciliation efficiency reached by $l$-DTE according to \Cref{eq:recon-eff-dir,eq:recon-eff-rev}, $l\in\qty{2,3,4}$ (black, red and blue plots, respectively), $\tilde{V}_m=1$ and $\xi = 0.02$. Solid and dashed lines correspond to the efficiency considering heterodyne and homodyne detection, respectively.}
    \end{figure}

    
    We plotted the reconciliation efficiencies of \Cref{eq:recon-eff-dir,eq:recon-eff-rev} in \Cref{fig:DTE-recon-eff} for heterodyne and homodyne detections with $\tilde{V}_m=1$, $\xi = 0.02$ and considering $l\in\qty{2,3,4}$ for binary expansion (corresponding to the black, red, and blue graphs, respectively). There are some interesting points to be highlighted. One is that a $l$-DTE-based reconciliation seems to have the same performance for RR and DR (in the applicable range of snr for both directions), which can imply a symmetry between $I(\mathcal{D}_i(X);Y)$ and $I(\mathcal{D}_i(Y);X)$. 
    
    Second, the maximum reconciliation efficiency appears as a decreasing function of the SNR. Although DTE does not perform well with homodyne-based CVQKD protocols, its usage should be restricted to protocols that use heterodyne measurements. In this case, a three-bit expansion is present for $\beta_{max}^\leftarrow>0.8$ for $SNR<0$ dB and $\beta_{max}^\leftarrow>0.9$ for $SNR<-3.6$dB. Here, another operational difference appears between SEC and DTE. In the SEC protocol, the subchannels with mutual information less than 0.02 bits (usually the first two bits in the sequence) are commonly disclosed, while in the DTE, even the fourth subchannel, which presents mutual information around 0.01 bits for $\opn{SNR}<-3.6\opn{dB}$, is crucial for the reconciliation efficiency to be greater than 0.9. The DTE-induced BIAWGN subchannels with $\opn{SNR}>-2 \opn{dB}$ have the first three bits above the 0.02 bit threshold commonly adopted for the SEC protocol.

    The difference in efficiencies of DTE reconciliation with homodyne and heterodyne detections is also notable; to discuss this, we consider the RR case. In the case of a homodyne protocol, Alice and Bob have correlated random Gaussian variables $X$ and $Y$ and by so, in \Cref{eq:recon-eff-rev}, $I(X;Y) = \log(1+\opn{SNR}_{hom})/2$ which gives
    \begin{equation}
    \beta^{hom}_{max} = \frac{H(\mathcal{D}(Y)) - l + \sum_{i=1}^{l}I(\mathcal{D}_i(Y);X)}{\log(1+\opn{SNR}_{hom})/2}.
    \end{equation}
    
    When heterodyne detection is used, both quadratures are homodyned and there are $2l$ binary sequences extracted using DET, $l$ for each quadrature. Due to the symmetry on the modulation and noise model, the $i$ -th subchannel from the $q$ and $p$ quadrature is statistically identical. Then, 
    \begin{equation}
    \beta^{het}_{max} = 2\cdot\frac{H(\mathcal{D}(Y)) - l + \sum_{i=1}^{l}I(\mathcal{D}_i(Y);X)}{\log(1+\opn{SNR}_{het})}.
    \end{equation}
    


\section{Conclusion}\label{sec:conclusion}
	
    We have presented an information reconciliation protocol designed for Continuous Variable QKD using the Distributional Transform, a tool from copula theory. Together with arguments from information theory, it was made possible to extract bit sequences from Gaussian random variables whose bits are undoubtedly independent. We showed that each bit in the binary expansion can be treated as an independent channel, and its capacities were estimated considering direct and reverse reconciliation for homodyne and heterodyne detection. We also derived the expressions for the reconciliation efficiency in both reconciliation directions and the results showed that maximum efficiency is reached in protocols with heterodyne detection and low SNR. More specifically, it is possible to reach $\beta_{max}^\leftarrow>0.9$ for $\opn{SNR}_{het}<-3.6\opn{dB}$ with a DTE of four bits. Future work could focus on the design of error-correcting codes for the DTE induced subchannels.








\bibliographystyle{IEEEtran}
\bibliography{2023-JCIS}

\begin{thebibliography}{10}
\providecommand{\url}[1]{#1}
\csname url@samestyle\endcsname
\providecommand{\newblock}{\relax}
\providecommand{\bibinfo}[2]{#2}
\providecommand{\BIBentrySTDinterwordspacing}{\spaceskip=0pt\relax}
\providecommand{\BIBentryALTinterwordstretchfactor}{4}
\providecommand{\BIBentryALTinterwordspacing}{\spaceskip=\fontdimen2\font plus
\BIBentryALTinterwordstretchfactor\fontdimen3\font minus
  \fontdimen4\font\relax}
\providecommand{\BIBforeignlanguage}[2]{{%
\expandafter\ifx\csname l@#1\endcsname\relax
\typeout{** WARNING: IEEEtran.bst: No hyphenation pattern has been}%
\typeout{** loaded for the language `#1'. Using the pattern for}%
\typeout{** the default language instead.}%
\else
\language=\csname l@#1\endcsname
\fi
#2}}
\providecommand{\BIBdecl}{\relax}
\BIBdecl

\bibitem{assche2006}
G.~V. Assche, \emph{Quantum {{Cryptography}} and {{Secret-Key
  Distillation}}}.\hskip 1em plus 0.5em minus 0.4em\relax {CAMBRIDGE UNIV PR},
  2006.

\bibitem{grosshans2002}
F.~Grosshans and P.~Grangier, ``Continuous {{Variable Quantum Cryptography
  Using Coherent States}},'' \emph{Phys. Rev. Lett.}, vol.~88, no.~5, p. 57902,
  Jan. 2002, doi: 10.1103/PhysRevLett.88.057902.

\bibitem{grosshans2003}
F.~Grosshans, G.~Van~Assche, J.~Wenger, R.~Brouri, N.~J. Cerf, and P.~Grangier,
  ``Quantum key distribution using gaussian-modulated coherent states,''
  \emph{Nature}, vol. 421, p. 238, 2003, doi: 10.1038/nature01289.

\bibitem{weedbrook2004}
C.~Weedbrook, A.~M. Lance, W.~P. Bowen, T.~Symul, T.~C. Ralph, and P.~K. Lam,
  ``Quantum cryptography without switching,'' \emph{Phys. Rev. Lett.}, 2004,
  doi: 10.1103/PhysRevLett.93.170504.

\bibitem{weedbrook2012}
C.~Weedbrook, S.~Pirandola, R.~{Garc{\'i}a-Patr{\'o}n}, N.~J. Cerf, T.~C.
  Ralph, J.~H. Shapiro, and S.~Lloyd, ``Gaussian quantum information,''
  \emph{Rev. Mod. Phys.}, vol.~84, no.~2, pp. 621--669, 2012, doi:
  10.1103/RevModPhys.84.621.

\bibitem{laudenbach2018}
F.~Laudenbach, C.~Pacher, C.-H.~F. Fung, A.~Poppe, M.~Peev, B.~Schrenk,
  M.~Hentschel, P.~Walther, and H.~H{\"u}bel, ``Continuous-{{Variable Quantum
  Key Distribution}} with {{Gaussian Modulation-The Theory}} of {{Practical
  Implementations}},'' \emph{Advanced Quantum Technologies}, vol.~1, no.~1, p.
  1800011, 2018, doi: 10.1002/qute.201800011.

\bibitem{jouguet2011}
P.~Jouguet, S.~{Kunz-Jacques}, and A.~Leverrier, ``Long-distance
  continuous-variable quantum key distribution with a {{Gaussian}}
  modulation,'' \emph{Phys. Rev. A}, vol.~84, no.~6, p. 62317, 2011, doi:
  10.1103/PhysRevA.84.062317.

\bibitem{jouguet2012}
P.~Jouguet, S.~{Kunz-Jacques}, E.~Diamanti, and A.~Leverrier, ``Analysis of
  imperfections in practical continuous-variable quantum key distribution,''
  \emph{Phys. Rev. A}, vol.~86, p. 32309, 2012, doi:
  10.1103/PhysRevA.86.032309.

\bibitem{brassard1994}
G.~Brassard and L.~Salvail, ``Secret-{{Key Reconciliation}} by {{Public
  Discussion}},'' in \emph{{{EUROCRYPT}}}.\hskip 1em plus 0.5em minus
  0.4em\relax {Berlin, Heidelberg}: {Springer Berlin Heidelberg}, 1994, pp.
  410--423, doi: 10.1007/3-540-48285-735.

\bibitem{yan2008}
H.~Yan, T.~Ren, X.~Peng, X.~Lin, W.~Jiang, T.~Liu, and H.~Guo, ``Information
  {{Reconciliation Protocol}} in {{Quantum Key Distribution System}},'' in
  \emph{{{ICNC}}}.\hskip 1em plus 0.5em minus 0.4em\relax {IEEE}, 2008,
  10.1109/icnc.2008.755.

\bibitem{nguyen2004}
K.-C. NGUYEN, G.~VAN~ASSCHE, and N.~J. CERF, ``Side-{{Information Coding}} with
  {{Turbo Codes}} and its {{Application}} to {{Quantum Key Distribution}},'' in
  \emph{International {{Symposium}} on {{Information Theory}} and Its
  {{Applications}}}, {Parma, Italy, October}, Oct. 2004.

\bibitem{qian2009}
X.~Qian, G.~He, and G.~Zeng, ``Realization of error correction and
  reconciliation of continuous quantum key distribution in detail,''
  \emph{Science in China Series F: Information Sciences}, vol.~52, no.~9, pp.
  1598--1604, Sep. 2009, doi: 10.1007/s11432-009-0147-0.

\bibitem{lu2010}
Z.~Lu, L.~Yu, K.~Li, B.~Liu, J.~Lin, R.~Jiao, and B.~Yang, ``Reverse
  reconciliation for continuous variable quantum key distribution,''
  \emph{Science China Physics, Mechanics and Astronomy}, vol.~53, no.~1, pp.
  100--105, Jan. 2010, doi: 10.1007/s11433-010-0069-2.

\bibitem{grosshans2003a}
F.~Grosshans, N.~J. Cerf, J.~Wenger, R.~{Tualle-Brouri}, and P.~Grangier,
  ``Virtual entanglement and reconciliation protocols for quantum cryptography
  with continuous variables,'' \emph{Quantum Info. Comput.}, vol.~3, no.~7, pp.
  535--552, Oct. 2003, doi: 10.5555/2011564.2011570.

\bibitem{assche2004}
G.~V. Assche, J.~Cardinal, and N.~J. Cerf, ``Reconciliation of a
  quantum-distributed {{Gaussian}} key,'' \emph{IEEE TIT}, vol.~50, no.~2, pp.
  394--400, 2004, doi: 10.1109/tit.2003.822618.

\bibitem{jouguet2014}
P.~Jouguet, D.~Elkouss, and S.~{Kunz-Jacques}, ``High-bit-rate
  continuous-variable quantum key distribution,'' \emph{Phys. Rev. A}, vol.~90,
  no.~4, p. 42329, 2014, doi: 10.1103/PhysRevA.90.042329.

\bibitem{mani2021}
H.~Mani, T.~Gehring, P.~Grabenweger, B.~{\"O}mer, C.~Pacher, and U.~L.
  Andersen, ``Multiedge-type low-density parity-check codes for
  continuous-variable quantum key distribution,'' \emph{Physical Review A},
  vol. 103, no.~6, p. 062419, Jun. 2021, doi: 10.1103/PhysRevA.103.062419.

\bibitem{bai2017}
Z.~Bai, S.~Yang, and Y.~Li, ``High-efficiency reconciliation for continuous
  variable quantum key distribution,'' \emph{Japanese Journal of Applied
  Physics}, vol.~56, no.~4, p. 44401, Mar. 2017, doi: 10.7567/jjap.56.044401.

\bibitem{leverrier2008}
A.~Leverrier, R.~All{\'e}aume, J.~Boutros, G.~Z{\'e}mor, and P.~Grangier,
  ``Multidimensional reconciliation for a continuous-variable quantum key
  distribution,'' \emph{Phys. Rev. A}, vol.~77, no.~4, p. 42325, 2008, doi:
  10.1103/PhysRevA.77.042325.

\bibitem{milicevic2018}
M.~Milicevic, C.~Feng, L.~M. Zhang, and P.~G. Gulak, ``Quasi-cyclic multi-edge
  {{LDPC}} codes for long-distance quantum cryptography,'' \emph{npj Quantum
  Information}, vol.~4, no.~1, p.~21, Dec. 2018, doi:
  10.1038/s41534-018-0070-6.

\bibitem{diamanti2016}
E.~Diamanti, H.-K. Lo, B.~Qi, and Z.~Yuan, ``Practical challenges in quantum
  key distribution,'' \emph{npj Quantum Information}, vol.~2, no.~1, p. 16025,
  Nov. 2016, doi: 10.1038/npjqi.2016.25.

\bibitem{araujo2018}
L.~Ara{\'u}jo, F.~Assis, and B.~Albert, ``Novo protocolo de reconcilia\c{c}\~ao
  de chaves secretas geradas quanticamente utilizando c\'odigos {{LDPC}} no
  sentido {{Slepian-Wolf}},'' in \emph{Anais de {{XXXVI Simp\'osio Brasileiro}}
  de {{Telecomunica\c{c}\~oes}} e {{Processamento}} de {{Sinais}}}.\hskip 1em
  plus 0.5em minus 0.4em\relax {Sociedade Brasileira de
  Telecomunica\c{c}\~oes}, 2018, doi: 10.14209/sbrt.2018.315.

\bibitem{bloch2006}
M.~Bloch, A.~Thangaraj, S.~McLaughlin, and J.-M. Merolla, ``{{LDPC-based
  Gaussian}} key reconciliation,'' in \emph{2006 {{IEEE Information Theory
  Workshop}}}.\hskip 1em plus 0.5em minus 0.4em\relax {Punta del Este,
  Uruguay}: {IEEE}, 2006, pp. 116--120, doi: 10.1109/ITW.2006.1633793.

\bibitem{lodewyck2007}
J.~Lodewyck, M.~Bloch, R.~{Garc{\'i}a-Patr{\'o}n}, S.~Fossier, E.~Karpov,
  E.~Diamanti, T.~Debuisschert, N.~J. Cerf, R.~{Tualle-Brouri}, S.~W.
  McLaughlin, and P.~Grangier, ``Quantum key distribution over 25 km with an
  all-fiber continuous-variable system,'' \emph{Phys. Rev. A}, vol.~76, no.~4,
  2007, doi: 10.1103/PhysRevA.76.042305.

\bibitem{sempi2016}
F.~Durante and C.~Sempi, \emph{Principles of Copula Theory}.\hskip 1em plus
  0.5em minus 0.4em\relax {Hoboken}: {CRC Press}, 2016,
  \url{http://site.ebrary.com/id/11074681}.

\bibitem{cover2006}
J.~A.~T. Thomas M.~Cover, \emph{Elements of {{Information Theory}}}.\hskip 1em
  plus 0.5em minus 0.4em\relax {Wiley John + Sons}, 2006.

\bibitem{nelsen2003}
R.~B. Nelsen, J.~J. {Quesada-Molina}, J.~A. {Rodr{\'{\i}}guez-Lallena}, and
  M.~{\'U}beda-Flores, ``Kendall distribution functions,'' \emph{Statistics \&
  Probability Letters}, vol.~65, no.~3, pp. 263--268, Nov. 2003, doi:
  10.1016/j.spl.2003.08.002.

\bibitem{nha2005}
H.~Nha and H.~J. Carmichael, ``Distinguishing two single-mode {{Gaussian}}
  states by homodyne detection: {{An}} information-theoretic approach,''
  \emph{Physical Review A}, vol.~71, no.~3, p. 032336, Mar. 2005, doi:
  10.1103/PhysRevA.71.032336.

\bibitem{steeg2016}
G.~V. Steeg, ``Gregversteeg/{{NPEET}}: {{Non-parametric}} entropy estimation
  toolbox,'' Nov. 2016, \url{https://github.com/gregversteeg/NPEET}.

\bibitem{kraskov2004}
A.~Kraskov, H.~St{\"o}gbauer, and P.~Grassberger, ``Estimating mutual
  information,'' \emph{Physical Review E}, vol.~69, no.~6, p. 066138, Jun.
  2004, doi: 10.1103/PhysRevE.69.066138.

\end{thebibliography}

\end{document}